\documentclass[aps,twocolumn,showpacs,nofootinbib,groupedaddress]{revtex4}
\usepackage{amsmath,amssymb,amsbsy}
\usepackage{graphicx}
\begin{document}
\title{Comment on ``First observation of doubly charmed baryon $\boldsymbol
\Xi_{\bf cc}^+$''}
\author{V.V.Kiselev}
\email{kiselev@th1.ihep.su}
\affiliation{State Research Center "Institute for High Energy Physics" \\
{Protvino, Moscow region, 142280 Russia}\\
Fax: +7-0967-744739}
\author{A.K.Likhoded,}
\affiliation{State Research Center "Institute for High Energy Physics" \\
{Protvino, Moscow region, 142280 Russia}\\
Fax: +7-0967-744739}
\begin{abstract}
We speculate on a possible interpretation of events selected by the
SELEX collaboration and stress an insufficient evidence for the observation of
doubly charmed baryon.
\end{abstract}


\maketitle

In the recent paper \cite{selex} the SELEX collaboration reported on the first
observation of doubly charmed baryon. In this comment we argue the following:
\begin{enumerate}
\item
Given the reported events would be caused by the doubly charmed baryons, the
observed particle would have extremely exotic characteristics as concerns for
the lifetime as well as the production rate and signature.
\item
The events observed by the SELEX collaboration as described in \cite{selex}
could be ordinary treated as associated pair production of charmed particles
with no involvement of doubly charmed baryons.
\end{enumerate}

In the analysis of SELEX data, the $\Lambda_c^+$ sample contains 1630 events
with the full reconstruction of $\Lambda_c^+\to pK^-\pi^+$ decay mode. It is
important that the products of decay were identified by use of the
Ring-Imaging Cherenkov detector (RICH), and the secondary vertex shifted
from the interaction point of incoming particle with the target was resolved
due to the silicon vertex detector. Further, the Cabbibo-allowed decay of
$\Xi_{cc}^+\to \Lambda_c^+K^-\pi^+$ was searched for, and, thus, one expected
the {\bf strong correlation} of charged kaon-sign in decays of $\Lambda_c^+$
and $\Xi_{cc}^+$. The collaboration chosen to investigate the position of
additional $K^-\pi^+$ vertex between the primary one and that of $\Lambda_c^+$.
This condition could remove a lot of events under interest, but the further
analysis in \cite{selex} shown that it is not so restrictive for the candidate
events. 

Let us assume that the SELEX collaboration can prove the observation of doubly
charmed baryon $\Xi_{cc}^+$ as claimed in \cite{selex}. Then the measurement of
$\Xi_{cc}^+$ lifetime in accordance with Table 1 in \cite{selex} shows the
preferable value of 0.012 ps with the upper limit of 0.033 ps at the 90\%
confidence level. This result is in a deep contradiction with the theoretical
predictions \cite{revqq,OPEQQq} based on the Operator Product Expansion
generalized to the systems composed of two heavy quarks and a single light
quark. The total width of the doubly charmed baryon up to subleading terms is
given by the sum of two basic terms. The first is the double total width of
free charmed quark, i.e. the spectator contribution corrected by the coupled
effects, which are mainly determined by the negative correction due to the time
dilution of heavy quark motion in the rest frame of baryon,
$$
\Gamma_c [\Xi_{cc}^+] \approx \Gamma_c^{\rm spect}\, \left(1-\frac{\langle
v_c^2 \rangle}{2}\right).
$$
The spectator width $\Gamma_c^{\rm spect}$ slightly depends on the
normalization scale coming from the higher-order corrections in QCD, so that it
is, with a good accuracy, independent of hadron state containing the charmed
quark. Therefore, the value of $\Gamma_c^{\rm spect}$ cannot be essentially
changed with respect to that of in $D$ mesons. The averaged square of charmed
quark velocity in the baryon can be estimated in the model-dependent way. Its
value is about 0.15, which gives a typical size of subleading corrections. The
second numerically-essential term in the total width of $\Xi_{cc}^+$ is
determined by the weak rescattering of charmed and $d$ quarks
$\Gamma_{WS}[\Xi_{cc}^+]$. Its value parametrically depends on the wave
function of baryon, which was in detail studied in Ref. \cite{revqq}. Thus, the
total width is equal to
$$
\Gamma_{\rm tot}[\Xi_{cc}^+] \approx 2\, \Gamma_{c}[\Xi_{cc}^+] +
\Gamma_{WS}[\Xi_{cc}^+],
$$
where the numerical contribution of weak scattering contribution is about 60\%.
Then the lifetime is predicted by the estimate
$$
\tau[\Xi_{cc}^+] = 0.16\pm 0.05\;{\rm ps},
$$
which is much greater than the measured value for the candidates reported by
the SELEX  collaboration. Note, that the extremely short lifetime would point
to the unexpected breaking of OPE in the doubly heavy baryons, since one should
essentially change the mass of charmed quark as well as drastically enlarge the
baryon wave function in contradiction with the description of hadrons
containing a single charmed quark\footnote{The very strict and pessimistic
approach to the accuracy of theoretical predictions for the lifetimes of
hadrons with the charmed quark gives the uncertainty about 50\%, but not the
factor of 20.}. We have to emphasize that the analysis of lifetime in Ref.
\cite{revqq} was performed in a consistent way with the data on the lifetimes
of baryons with the single charmed quark, i.e. $\Lambda_c$ and $\Xi_c^{+,0}$.
Thus, the lifetime of candidates observed by the SELEX collaboration is
extremely exotic for the doubly charmed baryon $\Xi_{cc}^+$. In addition, the
measured lifetime is even shorter than the lifetime of so-called ``sideband
region'' events (about 0.038 ps), and it is close to the single-event
resolution of 0.020 ps.

Next, the production rate of selected candidates if they are really the events
with the production of doubly charmed baryon, is extremely high. Indeed, the
SELEX collaboration found that about 20\% of $\Lambda_c^+$ events in its total
sample are produced by $\Xi_{cc}^+$ \cite{selex}. The mechanism for the
production of doubly charmed baryons in the strong interactions supposes the
production of two pairs $c\bar c$, i.e. four heavy quarks (see review in
\cite{revqq}). At high energies of parton subprocesses as the gluon fusion
$gg\to cc\bar c\bar c$ and the quark-antiquark annihilation $q\bar q\to cc\bar
c\bar c$, the hard production of heavy quarks is suppressed in comparison with
the production of single pair $c\bar c$ by the factor of $10^{-2}$, that is
smaller than the reported contents of doubly heavy baryons in the production of
$\Lambda_c^+$. This theoretical expectation is strongly convinced, since it is
based on the direct measurement of probability for the gluon splitting into the
pair of $c\bar c$ by the L3, ALEPH and OPAL collaborations in the
electron-positron annihilation at $Z$ boson peak, where one got $P(g\to c\bar
c) \sim 3\cdot 10^{-2}$ \cite{annihil}. In the process of gluon fusion dominant
at the SELEX energies, the total energy in the parton subprocess is essentially
less than the center-of-mass energy of hadron collisions because of the parton
luminosity. In this way, the hard production supposes a strong threshold effect
for four heavy quarks. This threshold suppression is significant for the
energies of SELEX operation, so that it results in the additional factor of
$10^{-2}$ \cite{hadr,prodQQq}. One should take into account the fact that the
hadronization of four heavy quarks results in a fraction of doubly heavy
baryons about 10\%, since an essential part of events with two charmed quarks
gives the production of two mesons or baryons each containing the single
charmed quark\footnote{The physical picture for such the suppression is quite
transparent: the hard production of charmed quarks takes place in the volume
about the Compton length cubed, while the baryon wave function determines the
size of doubly charmed diquark by a transfer momentum $p\sim m_c \cdot v$
with the relative velocity of charmed quark motion $v\ll 1$, so that the
probability of the baryon production is given by the ratio $p^3/m_c^3 \sim
v^3\ll 1$ in contrast to the continuum contribution formed by the hadrons
containing a single heavy quark.}. Thus, the hard production mechanism gives
the suppression of doubly heavy baryon production by the factor of $10^{-5}$ at
the energies of fixed target experiments\footnote{We stress that the
discrepancy of measured $\Xi_{cc}^+$ yield with the theoretical expectations
reaches the value about $10^{-4}$, while the result of BELLE \cite{Abe:2002rb}
mentioned in \cite{selex} as concerns for the production rate for $J/\psi c\bar
c$ disagrees within the factor of 10.}. Another possibility is the production
mechanism with the intrinsic charm contents in the initial hadrons
\cite{intrinsic}. In this case the threshold suppression is absent, but the
normalization of intrinsic structure functions is suppressed by the factor of
$10^{-2}$, so that the production rate for the doubly charmed baryons is about
$10^{-3}$ of the total charm rate. Anyway, the \mbox{SELEX} candidates for the
doubly charmed baryon $\Xi_{cc}^+$ have exotically high production rate in
comparison with theoretical expectations. Moreover, a low observed value of
mean transverse momentum for $\Lambda_c^+$ from $\Xi_{cc}^+$ points to the
preference of intrinsic charm mechanism, but with extremely high, and, hence,
unacceptable, normalization of charm distribution in the initial hadrons. As
for the signature of events with the $\Xi_{cc}^+$ candidates, the conservation
of flavor in the strong interactions supposes the associative production of
hadrons containing two anti-charmed quarks. These quarks should decay and
produce two additional vertices shifted from the primary one as well as result
in the additional charged multiplicity in the decays. The SELEX collaboration
did not report on the enhancement of charged multiplicity in the selected evens
or on the appearence of additional decay vertices. So, the question is where
are two anti-charmed quarks? Do they preferably disappear by decays in the
primary vertex with a low charged multiplicity? What is a probability of such
the conditions?

The main problem of the interpretation is that the particle identification in
the additional vertex of two charged particles was not possible, since the
momenta were insufficient in order to reach the RICH. In this case the analysis
loses the most strong evidence for the production of two charmed quarks in
contrast to the dominant process with the yield of $c\bar c$ pair. This main
process, then, can lead to the associative production of $\Lambda_c^+$ with the
neutral anti-charmed particle decaying to two charged tracks of opposite signs
with an unobserved neutral component lost by the silicon vertex detector as
well as in the system of magnets. For example, the branching fraction of $\bar
D^0\to K^+\pi^-\pi^0$ is equal to $13.9\pm 0.9$\% \cite{PDG}. In that case,
since the neutral pion is lost from the analysis, one cannot reconstruct $D^0$,
in part, its momentum. Therefore, one cannot draw a conclusion on the
Lorentz-factor of the charmed meson in order to make some claims on its decay
vertex\footnote{At low momenta, the decay vertex of charmed meson could be
rather close to the primary one.}. Thus, there is no evidence against the
ordinary treatment for the events reported by the \mbox{SELEX} collaboration in
\cite{selex} as the associative production\footnote{The associative production
of $\Lambda_c^+$ and $\bar\Xi_c^0$ decaying to $K^+\pi^-\bar\Xi^0$ is also
possible, while the neutral anti-baryon further decays to $\bar\Lambda\pi^0$.}
of $\Lambda_c^+$ and $\bar D^0$. Moreover, the appropriate assignment of
charged tracks leads to the $\Lambda_c^+K^+\pi^-$ mass distribution presented
in Fig. 2(b) of Ref. \cite{selex}, where we can see a rather smooth histogram,
which does not contain any significant peaks, but it exhibits a slow increase
of events in the mass region of $3.7-4.0$ GeV. This behaviour could be expected
if we suppose the associative production of $\Lambda_c^+$ with the charmed
particle, since in the case of full reconstruction one should observe an
ordinary threshold distribution starting at the energy $M[\Lambda_c^+]+M[\bar
D^0] \approx 4.1$ GeV, while the lose of neutral component in the decays of
$\bar D^0$ results in the smearing of threshold effect at lower
masses. 
{Unfortunately, the SELEX collaboration did not present a
comparison of mass distributions in the production of $\Lambda_c^+$ with the
expected form calculated with a Monte Carlo generator well describing the
events processed by the apparatus.} 
Next, the number of $\Lambda_c^+$ events
with the additional vertex 
is suppressed in comparison with the total rate of $\Lambda_c^+$.
{We do not find a direct claim on the number
of events with the vertex separation greater than a fixed cut-off, so that we
extract the amount of events under interest from the data on Fig. 2 in
\cite{selex}, where in
the region of $3.2-4.0$ GeV one can count several hundreds events.}
The value of suppression is
given by a typical efficiency for the reconstruction of additional vertex
(something about several per cents), only, while one should expect a stronger
suppression, because the production cross section for the doubly charmed
baryons has to be significantly less than the cross section for the inclusive
production of $\Lambda_c^+$. Finally, the wrong-assignment of kinematics can
result in the fake peaks shown in Fig. 2 (a) and (c) of \cite{selex}.

In conclusion, we show that the SELEX paper does not provide sufficient support
for its claim of evidence for the observation of doubly charmed baryon
$\Xi_{cc}^+$.

This work is supported in part by the Russian Foundation for Basic Research,
grants 01-02-99315, 01-02-16585, and 00-15-96645.


\end{document}